\documentclass[
 aip,
 amsmath,amssymb,
 reprint,floatfix,
]{revtex4-1}

\usepackage{graphicx}% Include figure files
\usepackage{dcolumn}% Align table columns on decimal point
\usepackage{bm}% bold math
\usepackage[utf8]{inputenc}
\usepackage[T1]{fontenc}
\usepackage{mathptmx}
\usepackage{comment}

\usepackage{ulem}
\usepackage{xcolor}

\usepackage{soul}
\begin{document}

\preprint{}

\title{Nonlinear Signal Distortion Corrections Through Quantum Sensing}

\author{Kevin R. Chaves}
\affiliation{Lawrence Livermore National Laboratory, Livermore, CA 94550, USA}
\author{Xian Wu}
\author{Yaniv J. Rosen}
\author{Jonathan L DuBois}

\date{\today}% It is always \today, today,

\begin{abstract}

Having accurate gate generation is essential for precise control of a quantum system. The generated gate usually suffers from linear and nonlinear distortion. Previous works have demonstrated how to use a qubit to correct linear frequency distortions but have not commented on how to handle nonlinear distortions. This is an important issue as we show that nonlinear amplitude distortions from the RF electronics can affect Rabi pulses by as much as 10\%. We present work that demonstrates how a transmon qubit can be used as a highly sensitive cryogenic detector to characterize these nonlinear amplitude distortions. We show that a correction can drive these errors down to <1\% over a 700 MHz range. This correction technique provides a method to minimize the effects of signal distortions and can be easily applied to broadband control pulses to produce higher fidelity arbitrary quantum gates.   
\end{abstract}

\maketitle

Achieving high fidelity quantum gates is dependent on the accurate control of signals sent to a quantum system\cite{doi.org/10.1038/nature08005, doi:10.1063/1.5089550}. In order to achieve this control, the effects from every stage of signal output must be known and compensated for through calibration procedures\cite{PhysRevLett.110.040502,PhysRevB.72.134519}.~When optimizing for a universal gate set, multiple methods of calibration have been proposed such as: optimizing gate fidelities found through benchmarking\cite{cerfontaine_self-consistent_2020, PhysRevLett.106.180504}, minimizing specific errors using phase estimation\cite{kimmel_robust_2015, doi.org/10.1016}, and modifying the hardware to reduce calibration burdens\cite{xu_high-fidelity_2020}.~In all these cases, it can save optimization time to first generate calibration tables for the electronics generating the signals and for the transfer functions of the wires leading down to the device.~This calibration is especially crucial when performing pulse shaping\cite{chow_optimized_2010, dds, PhysRevLett.103.110501} or optimizing arbitrary gates for applications like quantum simulation\cite{wu_high-fidelity_2020, Holland:2019zju} where the gates can be broadband.

Signal distortions are generally categorized into linear and nonlinear effects\cite{pozar2011microwave}. A linear distortion will multiply each frequency present in a signal by a constant value that depends on that frequency. Capacitors, inductors, and basic filters can all be described by linear ordinary differential equations, and fall into this category. There have been several proposals showing that the qubit itself can be used to accurately examine and correct for linear distortions\cite{arxiv,doi:10.1063/1.4813549,PhysRevLett.123.150501,doi:10.1038/nphys1994,PhysRevB.89.020503,PhysRevB.72.134519,doi:10.1063/1.5133894, arxiv, PhysRevLett.110.040502}. However, while these studies present useful techniques to characterize frequency distortions, they do not compensate for nonlinear distortions experienced by the qubit control pulse. 

In this letter, we demonstrate the presence of nonlinear distortions in the control signals, likely coming from limits on the hardware specifications of the Arbitrary Waveform Generator (AWG). In particular, we show that these distortions affect the amplitude of sinusoids coming out of an arbitrary waveform generator over a broadband range. The amplitude distortions cause >10\% error in the applied Rabi frequency, with several smaller distortions that we were unable to detect with room temperature electronics. We therefore analyzed the Rabi oscillations of a qubit to characterize these signal distortions, as they are highly sensitive to the qubit control pulse parameters. However, Rabi oscillations are insensitive to amplitude variations at pi multiples of the pulse. To account for this, we propose a detection scheme that sweeps both Rabi pulse amplitude and time to find signal distortions. Finally, we demonstrate a simple procedure that corrects for those distortions and returns Rabi frequencies with <1\% error. 

\begin{figure}[ht]
\includegraphics[width =  0.483\textwidth,height = 13 cm]{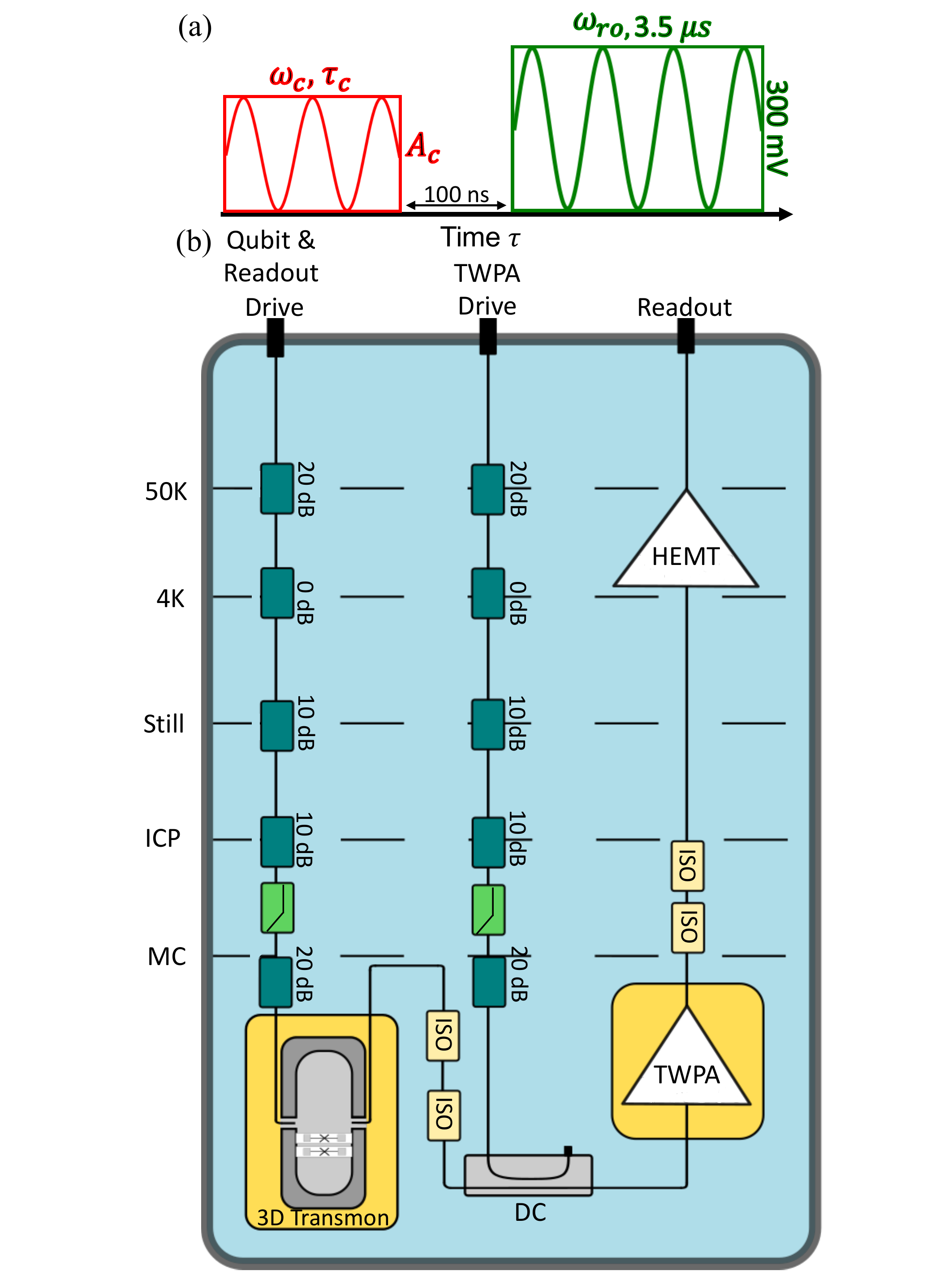}
\caption{The experimental setup for control pulse generation and calibration through qubit response. (a) A sketch of the control and readout pulses used in the experiment.  $\tau_c$, $A_c$, $\omega_c$/$\omega_{ro}$ indicates pulse duration, amplitude, and carrier frequency of the control pulse/ readout pulse. (b) A diagram of the experimental set up of cryogenic components inside a dilution refrigerator.}
\end{figure}

\begin{table}[ht]
\caption{Device Parameters.}
\label{tab:1}
\begin{ruledtabular}
\begin{tabular}{ccc}
 Name&Frequency& $T_1$ \\
\hline
Qubit 1& $4.09947$ $GHz$& 59.5 $\mu$s\\
Qubit 2&$4.80655$ $GHz$& 17.4 $\mu$s \\
Readout Mode&$7.07340$ $GHz$&
\end{tabular}
\end{ruledtabular}
\end{table}

\begin{figure}[ht!]
\includegraphics[width = 8.5cm,height = 7.5cm]{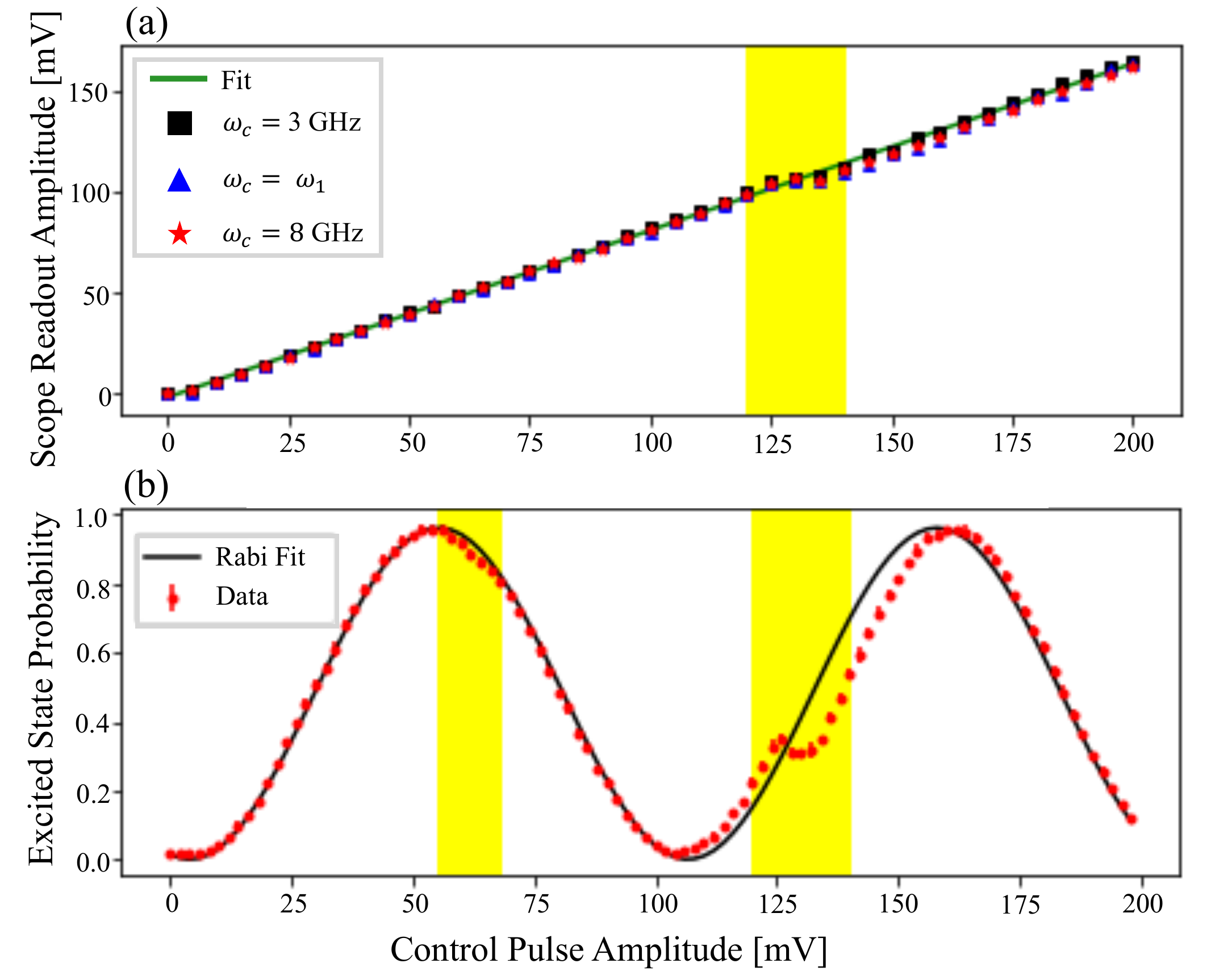}
\caption{
Measurement of amplitude distortions. (a) Direct oscilloscope measurements of $A_c$ generated by the AWG for $\omega_c$ = 3 GHz, $\omega_{1}$(Qubit 1), and 8 GHz. The three types of data points with error bars indicate the direct measurement data and the green linear fit shows the linear progression of the scope readout voltages before the amplitude distortion at 130 mV. (b) Excited state probability for Qubit 1 as a function of the control pulse amplitude. The red data points with error bars are measured excited state probability and the black line is the best fit according to the Rabi model in Eq.~\ref{eq:1}.}
\label{fig:2}
\end{figure}

The device used here is a 3D transmon composed of an aluminum cavity with a Si chip containing two transmon qubits mounted inside\cite {PhysRevLett.107.240501}.~Parameters for the cavity readout mode and two qubits are listed in Table~\ref{tab:1}.~The qubit drive and readout pulses are directly synthesized by a high speed AWG (Keysight M8195A). Both qubit drive and readout pulses are sinusoidal pulses with square envelopes as shown in Fig.\ref{fig:2}(a).~The generated pulses are then sent through coax cables, attenuators, and filters before reaching the input port of the 3D transmon. The transmitted signal coming out of the output port is sent through a pair of Isolators (ISO) and a Directional Coupler (DC) before being amplified by the Traveling Wave Parametric Amplifiers (TWPAs) and a High-Electron Mobility Transistor (HEMT) at the 4 K stage. The output signal is measured through a standard heterodyne detection scheme and the qubit's state is determined by the demodulated signals. The output signal has demodulation time on the order of $\mu$s to maintain good trade off between qubit decay and state distinguishability. 
The experimental setup is illustrated in Fig. \ref{fig:2}(b) with more details presented in Ref~13.

We first use room-temperature techniques to examine the AWG generated waveform.~A high speed oscilloscope\newline (Keysight MSO-X-91304A) is used to take direct measurements of the sinusoidal pulses. We sweep the pulse amplitude $A_c$ from 0 to 200 mV, and vary the carrier frequency $\omega_c$ from 3 to 8 GHz. Fig. \ref{fig:2}(a) shows the oscilloscope amplitude measurements for three carrier frequencies in the sweep as well as the expected linear output model (shown as green line). From the measurement result, we observe one obvious amplitude distortion in the highlighted region centered around $A_c$ = 130 mV for all three frequencies. In fact, this amplitude distortion is present throughout the carrier frequency sweep between 3 to 8 GHz. The distortion manifests as an amplitude dependent scale factor on the sinusoid, which preserves the signal's frequency distribution but alters the signal strength. We also note that the direct amplitude measurements for all three frequencies shown in Fig. \ref{fig:2}(a) significantly overlap, which suggests the amplitude distortion has minimal dependence on frequency.

Next, we apply the AWG generated sinusoidal pulses to the 3D transmon and use the qubit response as a direct measure of the pulse amplitude. Specifically, the carrier wave frequency is set on resonance of the qubit state transition frequency to drive Rabi oscillations. These oscillations depends on both the amplitude and duration of the pulse. For a two level system driven by a continuous sinusoidal pulse, the excited state population, $P_e$, follows the equation below\cite{book}:
\begin{equation}
\label{eq:1}
 P_e (\Omega_r,\tau_c) = sin^2 (\Omega_r\tau_c/2)
\end{equation}
where the Rabi frequency $\Omega_r \propto A_c$. If $\tau_c$ is fixed, we should expect $P_e$ to oscillate continuously as a function of $A_c$ according to Eq.~\ref{eq:1}. Fig. 2(b) shows a measurement of $P_e$ on Qubit 1 as a function of the control pulse amplitude with a 400ns pulse duration. The expected $P_e$ for non-distorted control pulses is plotted as the solid black curve. Our measurement shows a significant amplitude distortion around 130 mV, consistent with the oscilloscope results, and reveals another amplitude distortion around 65 mV, which is imperceptible in the oscilloscope results. Those signal distortions in the generated qubit control pulse could affect the gate fidelity and must be characterized and corrected. The qubit is more sensitive in detecting small changes or distortions in the generated signals compared to the standard techniques performed at room temperature. 

\begin{figure}[ht]
\includegraphics[width = 0.483\textwidth , height = 7.5cm]{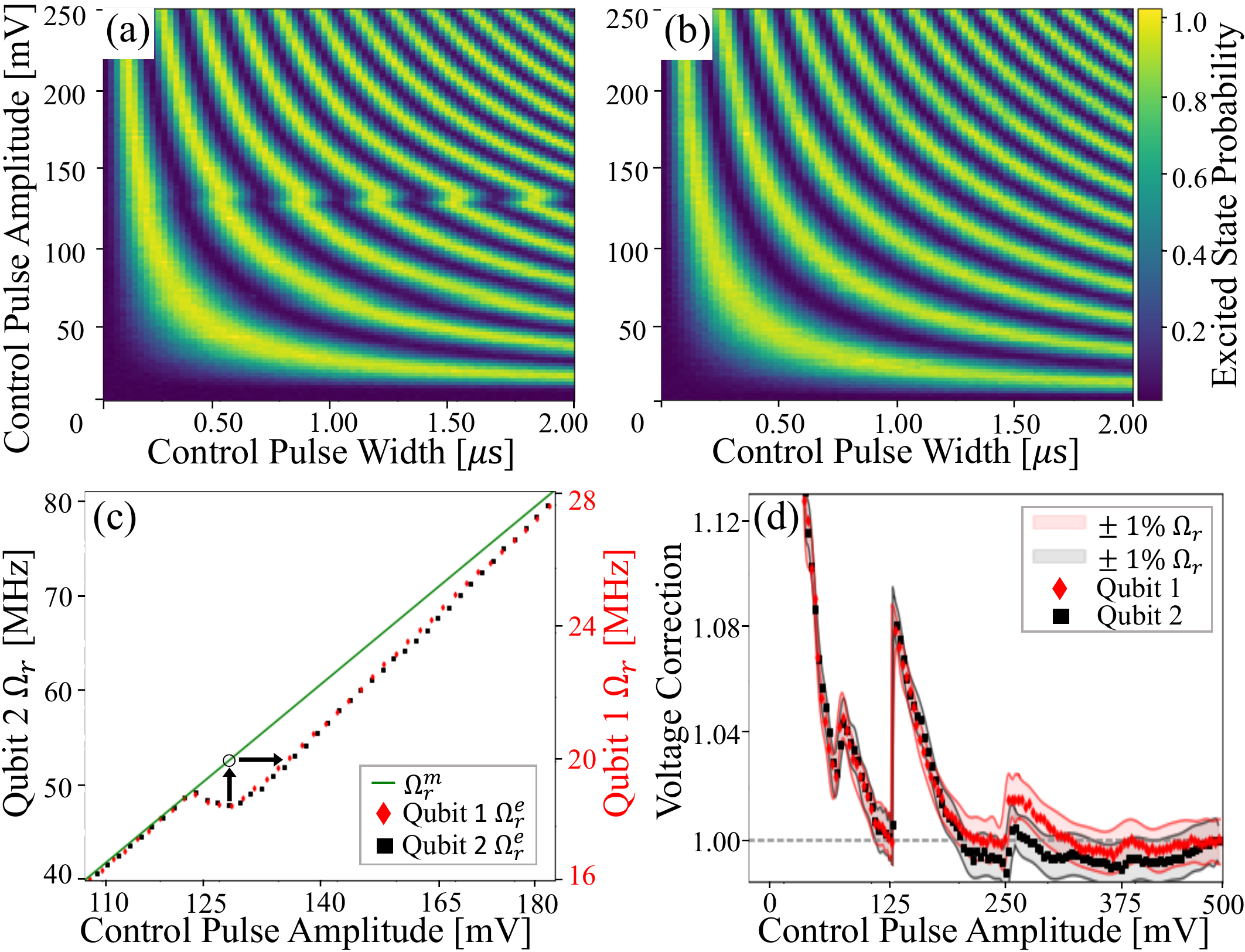}
\caption{
Measurement of amplitude distortion through the Rabi map experiment. (a),(b) Rabi oscillations of Qubit 1 for AWG amplitude range from 0 - 250~mV before and after voltage correction. 
(c) Experimentally extracted Rabi frequencies $\Omega_r^e$ for Qubit 1 (red diamonds) and Qubit 2 (black squares) for AWG output range between 110 - 180 mV. Sampled pulse amplitudes are in 2~mV step for Qubit 1 and 1.8~mV for Qubit 2. The modeled Rabi frequency $\Omega_r^m$ for linear amplitude output is indicated by the green solid line. 
(d) Voltage corrections for Qubit 1 (red diamonds) and Qubit 2 (black squares) over the full output range of AWG.  Shaded regions for both qubits indicates the voltage correction to cause $\pm 1\%$ error in induced Rabi frequency $\Omega_r$.
}
\label{fig:3} 
\end{figure}

We now describe our approach to characterize the full dynamic range of the AWG and develop a correction function for it. The experiment presented in Fig.~\ref{fig:2}(b) is not sensitive to amplitude distortion for the entire amplitude range, for example during the crests and troughs.~To address this issue, we conduct a Rabi map experiment which evaluates the qubit's Rabi oscillations as a function of pulse duration for the full dynamic range of the AWG. This experiment is done on both qubits in the 3D transmon, whose frequencies are 700 MHz apart. This helps determine if there is any frequency dependence on these amplitude distortions by characterizing these distortions at two points in frequency. Fig.~\ref{fig:3}(a) shows a segment of the Rabi map results, for $A_c$ from 0 to 250 mV on Qubit 1, which reveals several nonlinear distortions in the Rabi frequency.

To characterize the nonlinear distortions seen by both qubits, we extract the Rabi frequency $\Omega_r^e(A_c)$ from the Rabi map experiment in which the AWG amplitude is varied from 0~mV to 500~mV on Qubit 1 and 25~mV to 498~mV on Qubit 2. The extracted Rabi frequencies are shown as the red diamonds and the black squares, for Qubit 1 and Qubit 2 respectively, in Fig. 3(c). The input line attenuation varies across these two qubits, whose frequencies are about 700~MHz apart from each other. Therefore, the extracted Rabi frequencies for Qubit 1 and Qubit 2 are in different frequency ranges. Still, distortions for both qubits follow a similar pattern as the applied AWG amplitude varies. This implies that frequency dependence on the amplitude distortions are minimal in the qubit drive line. The amplitude range shown in Fig.3(c) highlights the predominant amplitude distortion at 130 mV, which is also shown in Fig~\ref{fig:2}. We also detect significant amplitude distortions at 65 mV and 250 mV from our Rabi map experiment. To quantify these amplitude distortion, we build a linear model $\Omega_r^m$ based on the measured Rabi frequency, shown as the green line in Fig.~\ref{fig:3}(c). The grey circle indicates the expected Rabi strength at 130 mV drive amplitude for linear AWG output.

Next, we discuss how we engineered an amplitude correction function for the full range of AWG output based on the extracted Rabi frequency $\Omega_r^e$ and the linear model $\Omega_r^m$.
\begin{enumerate}
    \item Fit the Rabi oscillations at each amplitude to extract the real $\Omega_r^e$ at each tested $A_c$.
     \item Perform linear interpolation on the extracted Rabi frequency data $\Omega_r^{e}(A_c)$ for both qubits.
    \item For each desired amplitude value $A_c$, determine its desired linear modeled Rabi frequency, $\Omega_r^m(A_c)$. This step is illustrated in Fig.~\ref{fig:3}(c), for the case of $A_c$ = 130 mV, as the black up arrow pointing to the circle on the green line. 
    \item Search for an applied control amplitude $A_o$ from the extracted Rabi frequency data that minimizes Eq.~\ref{eq:2} below:
    \begin{equation}
    \label{eq:2}
    \min_{A_o} [ \Omega_r^e(A_o) - \Omega_r^m(A_c)]^2
    \end{equation}
    This step is demonstrated by the black right arrow in Fig.~\ref{fig:3}(c)
    \item For each $A_c$, we define the voltage correction factor as $VC[A_c]=A_o/A_c$.
\end{enumerate}

Following these steps, we construct the voltage correction for both qubits over their respective amplitude ranges. Voltage corrections for both qubits are seen as the data points with error bars in Fig.~\ref{fig:3}(d).
\begin{figure}[ht]
\includegraphics[width = 8.5cm,height = 4.5cm]{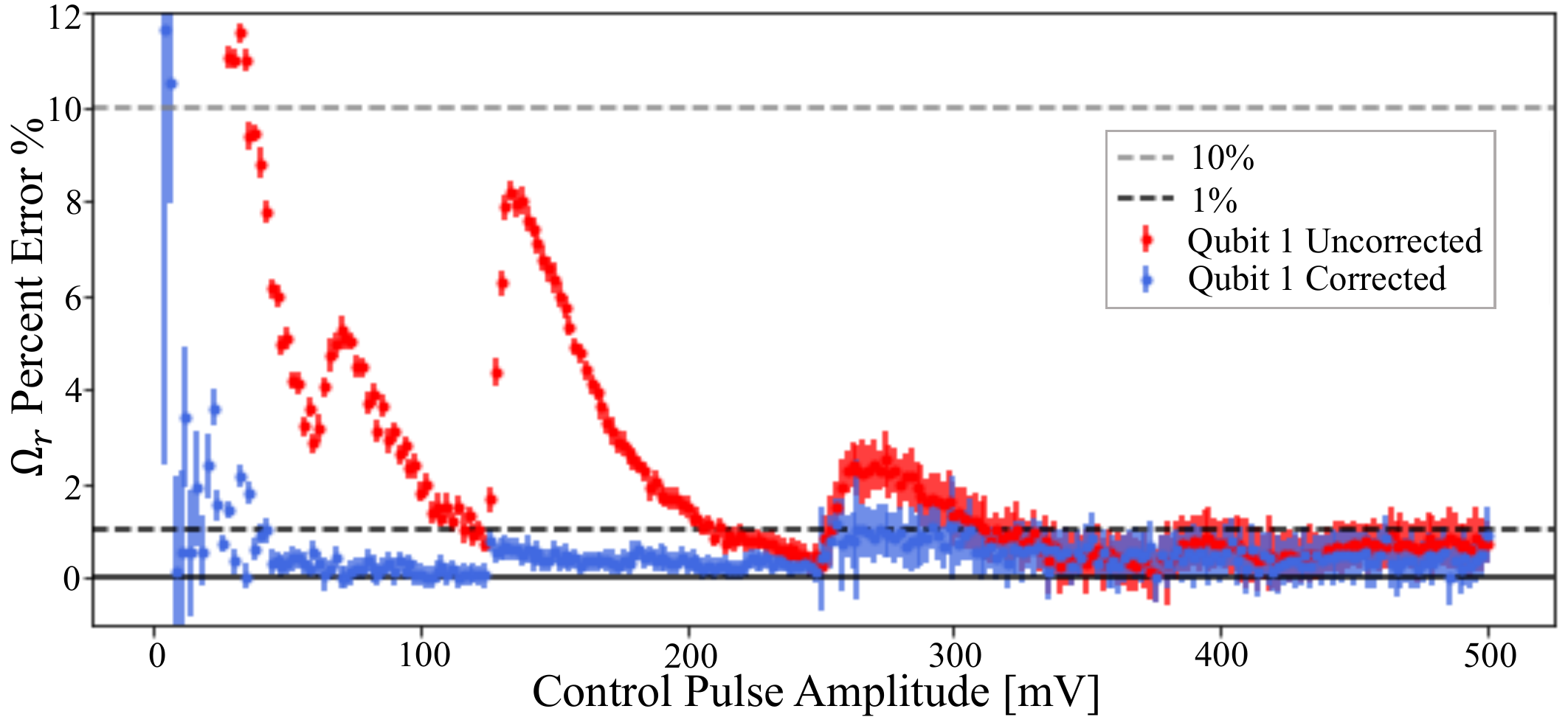}
\caption{\label{fig:4}The percent error in the Rabi frequency, $\Omega_r$, before and after correction is applied to each control pulse amplitude on Qubit 1.}
\end{figure}

We observe the extracted voltage corrections from both qubits follow the same pattern with small discrepancy, which could be due to either measurement error or minor frequency dependence in the amplitude distortions over this 700 MHz range. Since the deviations are minor, a single voltage correction function can be used over this broad frequency range. Next, we determine how much error this choice will impart on our system. We define an error range inside which any value of voltage correction would lead to a specified amount of error to the intended Rabi frequency $\Omega_r$. The shaded regions (red for Qubit 1, black for Qubit 2) shown in Fig.~\ref{fig:3}(d) indicates the error range corresponds to 1\% error in $\Omega_r$ for the entire output range. Continuous overlap of the two shaded regions for 1\% error in $\Omega_r$ is observed throughout the entire amplitude range. This indicates that a single voltage correction can be engineered for pulse correction over this 700 MHz range and generate <1\% error in the Rabi frequency.

To verify the effectiveness of this approach, we apply the generated voltage correction to the AWG output and repeat the Rabi map experiments with results shown in Fig.~\ref{fig:3}(b). We applied this voltage correction to the same qubits but during a different dilution refrigerator cool down period. The large nonlinear distortion centered around 130 mV has disappeared and other smaller distortions have been diminished as well. This indicates that the voltage correction can be applied with the same effectiveness between dilution refrigerator cool downs. Fig.~\ref{fig:4} shows the percent error in $\Omega_r$ for each $A_c$ applied to Qubit 1 both with and without correction. The percent error for the corrected $A_c$s falls below $1\%$ for most of the sweep and is always less than the error for the uncorrected $A_c$s. The percent error is larger for small amplitude values because the Rabi frequency is approaching decay rates ($T_1$) of the tested Qubit.

In summary, we demonstrate how a 3D transmon can be used as a highly sensitive cryogenic detector to characterize and correct for nonlinear amplitude distortions. By running Rabi map experiments on two different transmons over the complete AWG voltage range, we are able to characterize all the signal distortions in our quantum control hardware, revealing signal errors of >10\%. Using this characterization, we develop a correction function for our control pulses and successfully drive a qubit within the desired <1\% error threshold in Rabi frequency, for most control pulse amplitudes, in a 700 MHz range. This is a significant reduction in error compared to the ones generated by the uncorrected control pulse amplitudes in Fig. 3(a). This technique also provides the voltage correction factors needed to apply amplitude correction on broadband control pulses. By using this technique, in conjunction with others, gate errors caused by signal distortions can continue to be reduced and higher fidelity broadband gates can be achieved.

This work was performed under the auspices of the U.S. Department of Energy by Lawrence Livermore National Laboratory under Contract No. DE-AC52-07NA27344.  The authors gratefully acknowledge the DOE-ASCR Quantum Testbed Pathfinder Program under Award No. 2017-LLNL-SCW1631 for support of development of the quantum characterization and control correction methods described in this work and the NNSA-ASC Beyond Moores Law program (Grant No. LLNL-ABS-795437) for development and support of the LLNL quantum device integration testbed (QUDIT) facility. We thank James Colless, Irfan Sidiqqi, the Quantum Nanoelectronics Laboratory, and ARO/LPS for providing the transmon sample used in this work.
\bibliography{references}% Produces the bibliography via BibTeX.

\end{document}